\documentclass[a4paper,11pt]{article}
\pdfoutput=1
\usepackage{jcappub}
\usepackage[T1]{fontenc} 
\usepackage{graphicx,caption}
\usepackage{amsmath}
\usepackage{bm}
\usepackage{mathrsfs}
\usepackage{latexsym}
\usepackage{float}
\usepackage{color}
\usepackage[normalem]{ulem}
\usepackage{subcaption}
\usepackage{dcolumn}
\usepackage{multirow}
\usepackage[usenames,dvipsnames]{xcolor}
\usepackage{slashed}
\usepackage{booktabs}

\title{Investigating Dark Matter-Admixed Neutron Stars with NITR Equation of State in Light of PSR J0952-0607}
\author[a]{Pinku Routaray,}
\author[a]{Sailesh Ranjan Mohanty,}
\author[b]{H.C. Das,}
\author[a]{Sayantan Ghosh,}
\author[a]{P.J. Kalita,}
\author[c]{Vishal Parmar,}
\author[a,1]{Bharat Kumar \note{Corresponding author.}}
\affiliation[a]{Department of Physics \& Astronomy, National Institute of Technology, Rourkela 769008, India;}
\affiliation[b]{INFN Sezione di Catania, Dipartimento di Fisica,
Via S. Sofia 64, 95123 Catania, Italy;}
\affiliation[c]{School of Physics and Materials Science, Thapar Institute of Engineering and Technology, Patiala 147004, India.}
\emailAdd{routaraypinku@gmail.com}
\emailAdd{saileshranjanmohanty@gmail.com }
\emailAdd{harishdas.physics@gmail.com }
\emailAdd{sayantanghosh1999@gmail.com }
\emailAdd{probitjkalita@disroot.org}
\emailAdd{physics.vishal01@gmail.com}
\emailAdd{kumarbh@nitrkl.ac.in}

\abstract{
The {fastest and} heaviest pulsar, PSR J0952-0607, with a mass of $M=2.35\pm0.17 \ M_\odot$, has recently been discovered in the disk of the Milky Way Galaxy. In response to this discovery, a new RMF model, `NITR' has been developed. The NITR model's naturalness has been confirmed by assessing its validity for various finite nuclei and nuclear matter properties, including incompressibility, symmetry energy, and slope parameter values of 225.11, 31.69, and 43.86 MeV, respectively. These values satisfy the empirical/experimental limits currently available. The maximum mass and canonical radius of a neutron star (NS) calculated using the NITR model parameters are 2.355 $M_\odot$ and 13.13 km, respectively, which fall within the range of PSR J0952-0607 and the latest NICER limit. {This study aims to test the consistency of the NITR model by applying it to various systems. As a result, its validity is extensively calibrated, and all the nuclear matter and NS properties of the NITR model are compared with two established models such as IOPB-I and FSUGarnet. In addition, the NITR model equation of state (EOS) is employed to obtain the properties of a dark matter admixed NS (DMANS) using two approaches (I) single-fluid and (II) two-fluid approaches}. In both cases, the EOS becomes softer due to DM interactions, which reduces various macroscopic properties such as maximum mass, radius, tidal deformability, etc. {The various observational data such as NICER and HESS are used to constrain the amount of DM in both cases. Moreover, we discuss the impact of dark matter (DM) on the nonradial $f$-mode frequency of the NS in a single fluid case only and try to constrain the amount of DM using different theoretical limits available in the literature.}}
\begin{document}
\maketitle
\flushbottom
\section{Introduction}
\label{intro}
The neutron star (NS) is an extremely compact and complex object in the Universe, where extreme gravity meets extreme matter \cite{Huang_2022}. Astrophysicists are especially interested in determining its structure and dynamics, particularly after the detection of gravitational waves (GWs) from binary NS mergers \cite{Abbott_2017, Abbott_2018}, alongside the corresponding electromagnetic signals \cite{GW170819-EM-Drout_2017, GW170819-EM-Nicholl_2017, GW170819-EM-Chornock_2017, GW170819-EM-Cowperthwaite_2017}. Multifaceted data on NSs \cite{Balliet_2021} allows for the testing of various proposed EOSs, which are not feasible in terrestrial laboratories \cite{Feryal_2016}. The nuclear EOSs, which are critical in NS simulations, enable the understanding of several complex phenomena such as quark deconfinement ~\cite{Orsaria_2014}, phase transitions ~\cite{NKGfp_1992, Huang_2022}, very high magnetic fields \cite{Cardall_2001, Peng_2007, Vishal_2023}, appearances of hyperons \cite{NKGh_1985, NKGk1_1998, NKGk2_1999, Weissenborn_2012}, accretion of DM \cite{Sandin_2009, Kouvaris_2008, Das_2019, Das_2020, Das_2021}, pasta structures inside the crust \cite{Parmar_2022}, etc. with regards to the NS. Among various properties of the NS, its mass and radius are crucial in various astrophysical phenomena and their associated effects. Precise knowledge of these macroscopic properties of NSs enables the investigation of nuclear interactions in extreme environments.

Recently, a Galactic NS named PSR J0952-0607 has been detected in the disk of the Milky Way, which is the fastest and heaviest of its kind with a mass $M=2.35\pm0.17 \ M_\odot$ \cite{Romani_2022}. This discovery is in continuation of the pulsar PSR J0740+6620 ($M=2.08\pm0.07 \ M_\odot$ ~\cite{Cromartie_2020, Fonseca_2021}). Among several hundred available EOSs that are based on various forms of nuclear interactions such as {relativistic mean-field (RMF), Skyrme-Hartree-Fock (SHF)}, Gogny interaction, etc., only a few of them can reproduce the $M=2.35\pm0.17 \ M_\odot$ mass limit while satisfying nuclear matter constraints simultaneously \cite{Dutra_2012, Dutra_2014}. Therefore, it has become essential to revisit the optimization of the parameters of nuclear models to satisfy new constraints on the mass of the NS. 

In this article, we propose a new model named ``NITR" based on the extended RMF (E-RMF) framework, which is consistent with the underlying symmetry arising in Quantum Chromodynamics and solves the renormalization problem in conventional RMF theory \cite{Kumar_2017, Kumar_2018}. The E-RMF formalism has been successfully applied to various nuclear physics problems, including nuclear structure, reaction, and NS structure. Our proposed EOS reproduces the mass of the NS as $2.355 \ M_\odot$ and satisfies the nuclear matter constraints, making it a better alternative to the existing NS EOS. In this study, we use the NITR EOS to explore different properties of the NS with DM as an additional component inside it.

There is a plethora of evidence supporting the hypothesis that DM exists in the Universe, including rotation curves of galaxies, velocity dispersions, galaxy clusters, gravitational lensing, and cosmic microwave background \cite{Robert_DM_universe_2003}. Moreover, recent cosmological findings suggest that DM is not baryonic matter and must be a new type of matter that interacts only weakly with particles in the standard model \cite{LarsBergström_non-baryonic-DM_2000}. Despite substantial research into DM models from a particle physics perspective, the precise characteristics of DM particles are still unknown, although the weakly interacting massive particles (WIMPs) scenario has gained popularity. WIMPs can easily account for the known relic abundance of DM at the weak interaction scale and can be captured by NSs through elastic scattering with nucleons. The high density inside NSs means that the collisional energy loss of DM particles can be substantial, allowing them to be effectively captured inside such compact objects \cite{Kouvaris_2008, Kouvaris_2010, Goldman_1989, G_ver_2014, N-Raj_2018, Angeles_plb_2012}. The interaction of DM inside NSs has been studied using various models. In some models, DM interacts non-gravitationally with normal matter (NM) and is considered a single fluid system \cite{Grigorious_2017, Das_2019, harishjcap_2021, harishmnras_2021, harishprd_2021,pinku-prd_2023,Pinku_mnras_2023}. Other models ignore non-gravitational interaction between DM and NM, considering DM to interact through gravity, which results in a two-fluid system \cite{arpan_two-fluid_2022, Xiang_two-fluid_2014, Ellis_2018, Sandin_2009, Ciarcelluti_2011,rutherford_2022}.

In this work, using our new model `NITR', we calculate different properties of the NS such as mass, radius, tidal deformability, and nonradial $f$-mode frequency and compare with the very well-known models such as IOPB-I \cite{Kumar_2017} and FSUGarnet \cite{fsugarnet_2014}. To maintain consistency between the crust and the core, we calculate the unified EOS by employing the compressible liquid drop model \cite{Parmar_2022, Newton_2012} for the inner crust calculation and obtain the outer crust EOS using the most recent experimental masses and the Hartree-Fock-Bogoliubov-26 mass model \cite{Samyan_2002, Parmar_2022_1}. The NITR model predicts the maximum mass, canonical radius, and dimensionless tidal deformability are $M=2.355 M_\odot$, $R_{1.4}=13.13$ km, and $\Lambda_{1.4}= 682.84$ which closely satisfy the constraints by PSR J0952-0607, NICER$+$XMM (revised NICER) \cite{Miller_2021} and GW190814 \cite{Abbott_2020} respectively. But to produce other observable constraints such as $\Lambda_{1.4}$ by GW170817 \cite{Abbott_2017} and mass-radius constraints by  HESS 1731-347 \cite{HESS_2022}, we use a novel way by taking DM inside the NS and reproducing the different observational data without affecting the nuclear density functional. The dark matter admixed neutron star (DMANS) is investigated in two different approaches, (I) single-fluid and (II) two-fluid.

{In addition, we also constrain the amount of DM present inside the newly observed pulsar PSR J0740$+$6620 using the NITR model for both single and two-fluid approaches using various observational and theoretical bounds. In the single-fluid case, DM particles interact with ordinary matter through the exchange of standard model Higgs particles. The properties of DMANS is studied with varying the DM Fermi momentum ($k_f^{\rm DM}$) and the impact of DM on EOS, mass, radius, tidal deformability, and $f$-mode frequency is observed. Different mass-radius constraints are taken from NICER \cite{Miller_2019, Riley_2019}, revised NICER \cite{Miller_2021} as well as HESS J731$-$347 data \cite{HESS_2022} to put constraints on fraction of DM. Additionally, since the distribution of DM content within a NS affects the $f$-mode frequency of GW signals from mergers, we analyze the effects of DM on nonradial $f$-mode frequency. We attempt to constrain the amount of DM within the NS using the canonical nonradial $f$-mode frequency constraint proposed in different literature such as by Wen {\it et al.} \cite{Wen-fmode_2019}, Das {\it et al.} \cite{harishprd_2021}, Sotani {\it et al.} \cite{Sotani-kumar_2021}. On the other hand, in the two-fluid approach, it is considered that a Lagrangian with a single fermionic component, and the self-interaction of the fermionic DM is mediated by `dark scalar' and `dark vector' boson particles. The mass fractions of DM (ratio of DM mass to NS mass) as well as their coupling constants can be changed to observe their effects. Mainly, two different possibilities have been observed either DM halo or core depending on the mass and mass fractions and it is purely model-dependent \cite{arpan_two-fluid_2022,Xiang_two-fluid_2014,Michael_two-fluid_2022}. Finally, using various observational constraint, we try to constrain the percentage of DM in two fluid case.}
\section{RMF Formalism}
\label{sec:form}
The E-RMF formalism is based on the effective field theory-driven RMF model, which has resolved the renormalization issue in RMF theory while adhering to the fundamental QCD symmetries. This framework has been widely used in the past few years to study various nuclear physics issues, as demonstrated in Refs. \cite{Muller_1996, DelEstal_2001, Frun_1997, Singh_2013, Kumar_2017, Kumar_2018, Kumar_2020, Das_2020}. The E-RMF Lagrangian, which includes the fourth-order interaction between different mesons such as $\sigma$, $\omega$, $\rho$, and $\delta$ is given by \cite{FURNSTAHL_1996, Frun_1997, singh_2014, Kumar_2017, Kumar_2018}.
\begin{eqnarray}
{\cal L}_{\rm nucl.} & = &  \sum_{\alpha=p,n} \bar\psi_{\alpha}
\Bigg\{\gamma_{\mu}\bigg(i\partial^{\mu}-g_{\omega}\omega^{\mu}-\frac{1}{2}g_{\rho}\vec{\tau}_{\alpha}\!\cdot\!\vec{\rho}^{\,\mu}\bigg)-\bigg(M_{\rm nucl.} 
-g_{\sigma}\sigma\bigg)\Bigg\} \psi_{\alpha}
\nonumber \\
&&  
+\frac{1}{2}\partial^{\mu}\sigma\,\partial_{\mu}\sigma-\frac{1}{2}m_{\sigma}^{2}\sigma^2+\frac{\zeta_0}{4!}g_\omega^2(\omega^{\mu}\omega_{\mu})^2-\frac{\kappa_3}{3!}\frac{g_{\sigma}m_{\sigma}^2\sigma^3}{M_{nucl.}}-\frac{\kappa_4}{4!}\frac{g_{\sigma}^2m_{\sigma}^2\sigma^4}{M_{nucl.}^2}
\nonumber\\
&&
+\frac{1}{2}m_{\omega}^{2}\omega^{\mu}\omega_{\mu}-\frac{1}{4}W^{\mu\nu}W_{\mu\nu}
+\frac{1}{2}m_{\rho}^{2}\bigg(\vec\rho^{\mu}\!\cdot\!\vec\rho_{\mu}\bigg)-\frac{1}{4}\vec R^{\mu\nu}\!\cdot\!\vec R_{\mu\nu}
\nonumber\\
&&
-\Lambda_{\omega}g_{\omega}^2g_{\rho}^2\big(\omega^{\mu}\omega_{\mu}\big)\big(\vec\rho^{\,\mu}\!\cdot\!\vec\rho_{\mu}\big)\, .
\label{RMF}
\end{eqnarray}
Here, the relevant coupling constants are $g_\sigma$, $g_\omega$ and $g_\rho$, while the corresponding masses are $m_\sigma, m_\omega, m_\rho$ for $\sigma$, $\omega$, and $\rho$ mesons, respectively. The spatial component of the energy-momentum tensor provides pressure, and its zeroth component gives the energy density of the system. The details can be found in Ref. \cite{Kumar_2017, Kumar_2018, Serot_1986, Furn_1987, Reinhard_1988}. Using the energy density and pressure for the systems, one can estimate the different properties of the finite nuclei, nuclear matter, and NS, as discussed in Sec. \ref{RD}.

The Lagrangian density parameters given by Eq. (\ref{RMF}) have been calibrated using the simulated annealing method to optimize the Lagrangian density for a specified parameter space. The parameterization process has been elaborately explained in Refs. \cite{BKAgrawal_2005,BKAgrawal_2006,Kumar_2017,Kumar_2018}. We used experimental data for binding energies and charge radii of several nuclei, including $^{16}$O, $^{40}$Ca, $^{48}$Ca, $^{68}$Ni, $^{90}$Zr, $^{100,132}$Sn, and $^{208}$Pb, to fit the coupling constants or parameters and constrain the values of the nuclear matter incompressibility (K) and symmetry energy coefficient ($J$) within 210$-$245 MeV and 28$-$35 MeV, respectively. Additionally, to ensure that the maximum NS mass is around 2.35 $M_\odot$, we allowed the parameter $\zeta_0$ corresponding to the self-coupling of $\omega$ mesons to vary between 1.0 and 1.5. The obtained parameter set is given in Table \ref{tab:parameter}.

\begin{table}
\caption{{Masses of the $\sigma$, $\omega$, and $\rho$ mesons, the coupling constants for nucleons with mesons are tabulated for the IOPB-I \cite{Kumar_2018}, FSUGarnet \cite{fsugarnet_2014}, and NITR model. The mass of the nucleons ($M_{\rm nucl.}$) is taken as 939 MeV. Some of the nuclear matter properties are also given in the lower panel. The emperical/experimental values are given in the last column with their Refs. [a]\cite{Bethe_1971}, [b]\cite{Colo_2014}, [c]\cite{Danielewicz_2014}.}}
\scalebox{1.3}{
\begin{tabular}{ccccccccccc}
\hline
\hline
\multicolumn{1}{c}{}
&\multicolumn{1}{c}{IOPB-I}
&\multicolumn{1}{c}{FSUGarnet}
&\multicolumn{1}{c}{NITR}
&\multicolumn{1}{c}{Emp./Expt.}\\
\hline
$m_{\sigma}$  &  500.512 & 496.939 & 492.988 & \\
$m_{\omega}$  &  782.5  &  782.5 & 782.5 & \\
$m_{\rho}$  &  763.0 & 763.0 & 763.0 & \\
$g_{\sigma}$  &  10.392 &  10.504 & 9.869 & \\
$g_{\omega}$  &  13.345 &  13.700 & 12.682 & \\
$g_{\rho}$  &  11.121  &  13.889 & 14.192 & \\
$k_{3} $   &  1.496 &  1.367 & 1.724 & \\
$k_{4}$  &  -2.932  &  -1.399 & -5.764 & \\
$\zeta_{0}$  &  3.103 &  4.410  & 1.189 & \\
$\Lambda_{\omega}$  &  0.024 &  0.043 & 0.045 & \\
\hline
$\rho_0(\rm fm^{-3})$  & 0.149 & 0.153  & 0.155 & $0.148-0.185^{[a]}$\\
$\mathcal{E}_0{\rm (MeV)}$  & -16.10 & -16.23 & -16.34 & $-15 - - 17^{[a]}$\\
$K {\rm (MeV)}$   & 222.65 & 229.5 &  225.11 & $220-260^{[b]}$\\
$J {\rm (MeV)}$   & 33.30 & 30.95 &  31.69 & $30.20-33.70^{[c]}$\\
$L {\rm (MeV)}$   & 63.58 & 51.04 & 43.86 & $35.00-70.00^{[c]}$\\
\hline
\hline
\end{tabular}}
\label{tab:parameter}
\end{table}
\section{Unified EOS }
\label{ns}
To construct the unified EOS for the NS, we compute the EOS for both the crust and core by utilizing the effective field theory-driven RMF model and the NITR parameter set. The core EOS is obtained by imposing both the $\beta$-equilibrium and charge neutrality conditions on top of the E-RMF Lagrangian density (in Eq. \ref{RMF}). For the crust EOS, we adopt the methodology given in Ref. \cite{Parmar_2022_1, Parmar_2022, Carreau_2019, Carreau_2020}. To ensure the uniformity and more accurate determination of the properties of the NS, a unified EOS is required \cite{Pearson_2018, Parmar_2022}. The outer crust, which is arranged into different layers of nuclei, is determined using the most recent experimental masses \cite{Parmar_2022_1}. For the inner crust, we assume the Wigner-Seitz approximation, considering nuclear clusters in a body-centered cubic lattice structure employing the compressible liquid drop model approach, and calculate the energy of the cluster and surrounding neutron gas using the NITR EOS. The transition from the heterogeneous inner crust to a homogeneous core occurs at a density of $0.10115$ fm$^{-3}$ and a pressure of $0.41340$ MeV fm$^{-3}$ for the NITR EOS, which is consistent with the available constraints on NS crust \cite{Balliet_2021, newton_2021}. All the different parameters of NITR model is compared with well known  IOPB-I \cite{Kumar_2018} and FSUGarnet \cite{fsugarnet_2014} model and enumerated in Table \ref{tab:parameter}.

\begin{figure}[h]
   \centering
    \includegraphics[width = 0.5\textwidth]{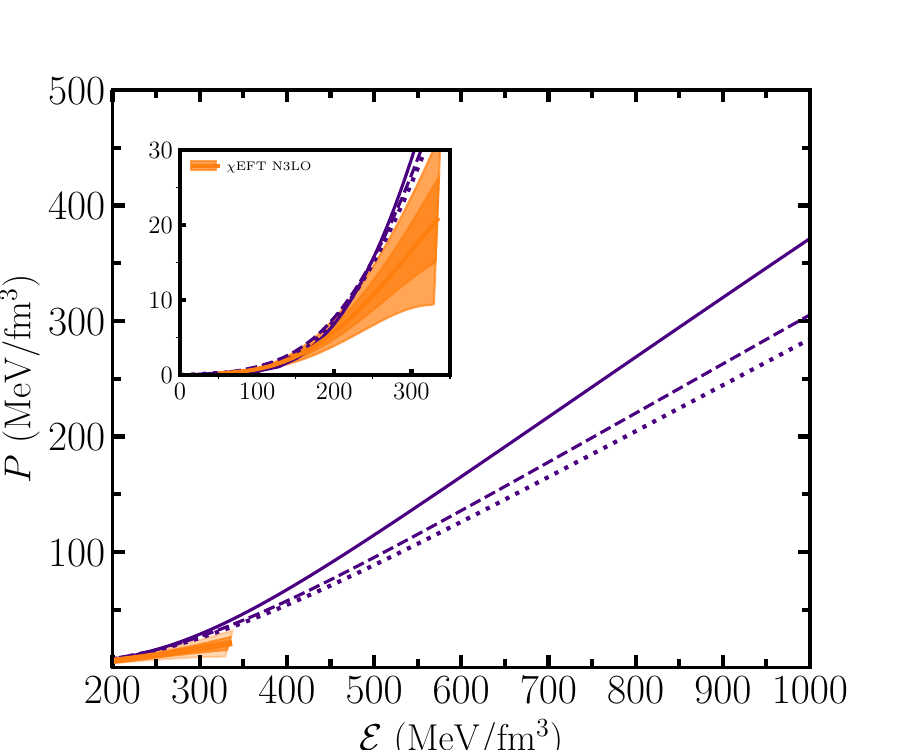}
    \includegraphics[width = 0.47\textwidth]{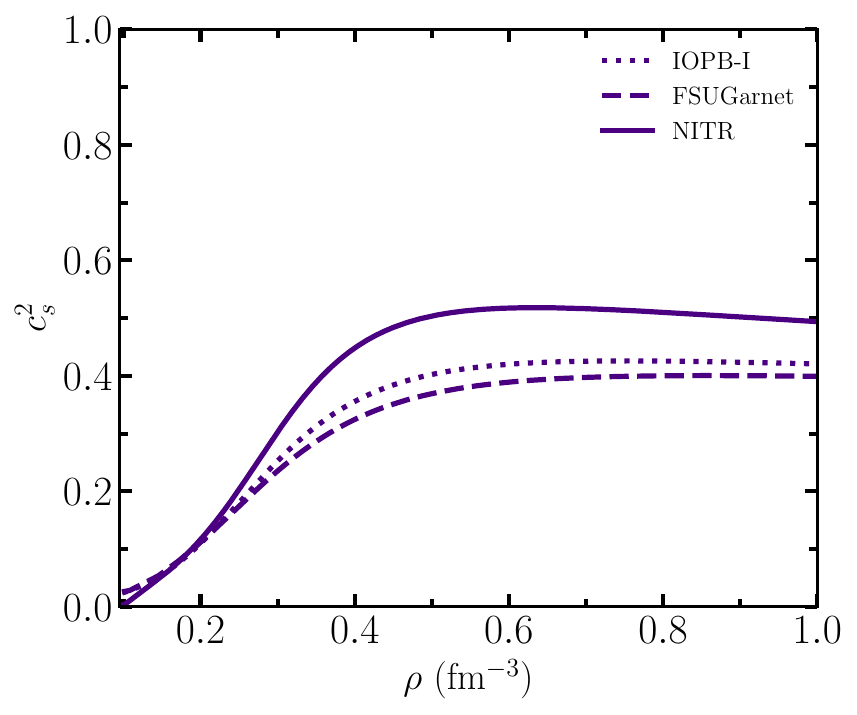}
    \caption{{{\it Left:} Unified EOSs for the NS for three parameter sets such as IOPB-I, FSUGarnet, and NITR EOSs. The orange shaded regions for the chiral EFT bounds \cite{Drischler_2021}. {\it Right:} The sound speed for considered EOSs.}}
    \label{fig:ns-eos}
\end{figure}

{In Fig. \ref{fig:ns-eos}, the left panel illustrates the unified EOS for NITR and two other EOSs IOPB-I and FSUGarnet, while the right panel depicts the corresponding speed of sound. Notably, NITR exhibits a stiffer behavior compared to IOPB-I and FSUGarnet. In the lower density regime, all three EOSs predict nearly identical pressures, resulting in similar canonical radii. However, as density increases, the pressure becomes stiffer, significantly influencing the maximum mass and its corresponding radius. Among these EOSs, NITR achieves the highest speed of sound value, as its pressure can support more mass compared to the other EOSs. Nonetheless, none of the EOSs violate the causality limit ($c_s^2=1$).}

\section{DM inside the NS}
During the evolution period of compact objects, such as NS, there is a certain probability that some amount of DM is captured inside it due to its high baryon abundance and huge gravitational potential. The accreted DM particles lose enough amount of energy due to their interactions with nucleons \cite{Das_2019, Kouvaris_2011, Goldman_1989}. Non-annihilating WIMPs were selected as the most popular DM candidate because WIMPs are the most generous type of DM particle and the thermodynamic residue. Other processes, such as the conversion of neutrons to scalar DM and the generation of scalar DM via bremsstrahlung, contribute to an increase in the abundance of DM contained within the NS \cite{Das_2019, Ellis_2018, Ellis_plb_2018}. The amount of DM mainly affects both the evolutionary history and the properties of the NS. There are two types of scenarios for DMANS: the single-fluid model and the two-fluid model, which we will discuss in the following subsections.
\subsection{Single fluid dark matter (SFDM) Model} 
\label{formalism-SF}
In the single-fluid model, DM particles interact with baryons by exchanging standard model Higgs. The structure of the interacting Lagrangian can be deduced as \cite{Grigorious_2017, Das_2019, harishjcap_2021, harishmnras_2021, harishprd_2021,pinku-prd_2023}:
\begin{eqnarray}
{\cal{L}}_{\rm SFDM} & = & \bar \chi \left[ i \gamma^\mu \partial\mu - M_\chi + y h \right] \chi + \frac{1}{2}\partial_\mu h \partial^\mu h - \frac{1}{2} M_h^2 h^2 + f \frac{M_{\rm nucl.}}{v} \bar \varphi h \varphi , 
\label{eq:SFLDM}
\end{eqnarray}
where, $\varphi$ and $\chi$ represent the nucleon and DM wave functions, respectively, while $h$ is the Higgs field. The mass of the DM (Neutralino) is assumed to be 200 GeV, and the Higgs mass ($M_h$) is 125 GeV. Since Neutralino is a supersymmetric particle, different coupling parameters are found in the electroweak sector of the standard model in the Higgs portal case, resulting in values of the DM-Higgs coupling ($y$) in the range of 0.001-0.1 \cite{MARTIN_1998}. However, a value of $y = 0.07$ is chosen in this study. The Yukawa coupling between Higgs and nucleon is $f M_{\rm nucl.}/v$, where $f$ is the proton-Higgs form factor, and $v$ is the vacuum expectation value assumed to be 0.35 \cite{Cline_2013} and 246 GeV \cite{harishmnras_2021, harishprd_2021, Das_2019}, respectively. The main assumption in this model is that the DM density ($\rho_{DM}$) is $10^3$ times less than the nucleon baryon density \cite{harishmnras_2021, Grigorious_2017, Das_2019}. One can calculate the Fermi momentum using the relation $k_f^{\rm DM} = (3\pi^2\rho_{DM})^{1/3}$. The value of $k_f^{\rm DM}$ is obtained to be $\sim 0.033$ GeV, but in the current work, it is varied between $0.00-0.05$ GeV to calculate the DM effects on various NS properties.

One can get the energy density and pressure by solving the Lagrangian density (in Eq. \ref{eq:SFLDM}) as in the following \cite{harishmnras_2021, Grigorious_2017, harishprd_2021}, 
\begin{eqnarray}
{\cal{E}}_{\rm SFDM} = \frac{2}{(2\pi)^{3}}\int_0^{k_f^{\rm DM}} d^{3}k \sqrt{k^2 + (M_\chi^\star)^2 } + \frac{1}{2}M_h^2 h_0^2 \ ,
\label{eq:edm}
\end{eqnarray}
\begin{eqnarray}
P_{\rm SFDM} = \frac{2}{3(2\pi)^{3}}\int_0^{k_f^{\rm DM}} \frac{d^{3}k \hspace{1mm}k^2} {\sqrt{k^2 + (M_\chi^\star)^2}} - \frac{1}{2}M_h^2 h_0^2 \,
\label{eq:pdm}
\end{eqnarray} 
where $M_\chi^* (=M_\chi-yh_0)$ is called the effective mass of the DM. Now, the total energy density and pressure for DMANS can be written as,
\begin{eqnarray}
{\cal{E}}={\cal{E}}_{\rm NS}+ {\cal{E}}_{\rm SFDM} \,, 
\nonumber
\\
{\rm and}
\hspace{1cm}
P=P_{\rm NS} + P_{\rm SFDM} \, .
\label{eq:EOS_total}
\end{eqnarray}

{To calculate the properties of DMANS in a single fluid approach, one can use the EOS in Eq. \ref{eq:EOS_total}. Here we have used the TOV equations \cite{Tolman_1939, Oppenheimer_1939} to calculate the mass and radius of the NS. Tidal deformability can be computed by solving its corresponding differential equation along with the TOV equation and the detailed theoretical formalism can be found in Refs.\cite{Hinderer_2008, KumarTide_2017}. The nonradial $f$-mode frequency can be determined using the Cowling approximation \cite{sotani-cowling_2011, Flores-cowling_2014, Ranea-Sandoval-cowling_2018, athul-fmode_2022}.}
\subsection{Two fluid dark matter (TFDM) Model}
The DM particles in the two-fluid DM (TFDM) model mainly self-interact through gravity, and the nature of the DM candidate can either be fermionic or bosonic. Here, we consider DM as a fermion and represent it by the field $\psi_D$. The interactions between the dark scalar meson ($\phi_D$) and the dark vector meson ($V_D^{\mu}$) with fermionic DM are represented by $g_{\rm sd}\Bar{\psi}\psi_D\phi_D$ and $g_{\rm vd}\Bar{\psi}\gamma_{\mu}\psi_DV_D^{\mu}$, where $g_{\rm sd}$ and $g_{\rm vd}$ are the coupling constants corresponding to the dark scalar and dark vector mesons, respectively. Similar to the baryons, one can define the effective potential for DM with exchanging bosons \cite{Fetter_1971}
\begin{eqnarray}
     V_{eff}(r)=\frac{g_{\rm vd}^{2}}{4\pi} \frac{e^{-m_{\rm vd}r}}{r}-\frac{g_{\rm sd}^{2}}{4\pi} \frac{e^{-m_{\rm sd}r}}{r}.
\label{eq:potential}
\end{eqnarray}
Now, one can write the Lagrangian density for the DM model, which is given by \cite{arpan_two-fluid_2022, Xiang_two-fluid_2014}
\begin{align}
{\cal L_{\rm TFDM}}=&\Bar{\psi_D}[\gamma_\mu(i\partial^\mu-g_{\rm vd}V^\mu)-(M_D-g_{\rm sd}\phi_D)]\psi_D
\nonumber \\ 
&
+\frac{1}{2}[\partial_\mu\phi_D\partial^\mu\phi_D-m_{\rm sd}^2\phi_D^2] -\frac{1}{4}V_{\mu\nu,D}V_D^{\mu\nu}+\frac{1}{2}m_{\rm vd}^2V_{\mu,D} V_D^\mu \, ,
\label{eq:TFLDM}
\end{align}
where, the mass of the DM candidate is represented by $M_D$, while $m_{sd}$ and $m_{vd}$ are the masses corresponding to the dark scalar meson and dark vector meson, respectively. Additionally, we define the term $V_{\mu \nu,D}$ as $\partial_\mu V_{\nu,D}-\partial_\nu V_{\mu,D}$.

In Section \ref{sec:form}, we introduced the E-RMF model, and in Subsection \ref{formalism-SF}, we discussed the DM admixed single-fluid model. Similar to these models, the mean-field approximation can be applied to the TFDM Lagrangian density given in Equation \ref{eq:TFLDM}. This allows us to obtain the energy density and pressure of the TFDM model, denoted by ${\cal E}_{\rm TFDM}$ and $P_{\rm TFDM}$, respectively. The expressions for ${\cal E}_{\rm TFDM}$ and $P_{\rm TFDM}$ can be found in the literature \cite{Xiang_two-fluid_2014,arpan_two-fluid_2022}
\begin{eqnarray}
{\cal E}_{\rm TFDM}=\frac{1}{\pi^2}\int_{0}^{k_D} dk \ k^2(k^2+M_D^{* 2})^{1/2} + \frac{g_{\rm vd}^2}{2m_{\rm vd}^2}\rho_D^2+\frac{m_{\rm sd}^2}{2g_{\rm sd}^2}(M_D-M_D^*)^2 \, ,
\label{TFE}
\end{eqnarray}
\begin{eqnarray}
P_{\rm TFDM} =\frac{1}{3}\frac{1}{\pi^2}\int_{0}^{k_D} dk \ \frac{k^4}{(k^2+M_D^{*2})^{1/2}} + \frac{g_{\rm vd}^2}{2m_{\rm vd}^2}\rho_D^2-\frac{m_{\rm sd}^2}{2g_{\rm sd}^2}(M_D-M_D^*)^2 \,.
\label{TFP}
\end{eqnarray}
The DM number density is denoted by $\rho_D$, while $M_D^*=M_D-g_{\rm sd}\phi_D$ is referred to as the effective mass of the DM candidate. The attractive potential $C_{\rm sd}=g_{\rm sd}/m_{\rm sd}$ and repulsive potential $C_{\rm vd}=g_{vd}/m_{vd}$ are both defined in units of GeV$^{-1}$. These potentials play a crucial role in the two-fluid DM (TFDM) model, where DM particles mainly self-interact through gravity. Specifically, they determine the nature of the interactions between DM and the dark scalar meson ($\phi_D$) and dark vector meson ($V_D^{\mu}$).
\section{Results and Discussions}
\label{RD}
In this section, we present our numerical results for NS. Various properties obtained with the newly developed parameter set NITR for DMANS in single fluid as well as two fluid approaches are also discussed.
\subsection{Single fluid dark matter}
\label{RD-SF}
\begin{figure}
   \centering
   \includegraphics[width = 0.6\textwidth]{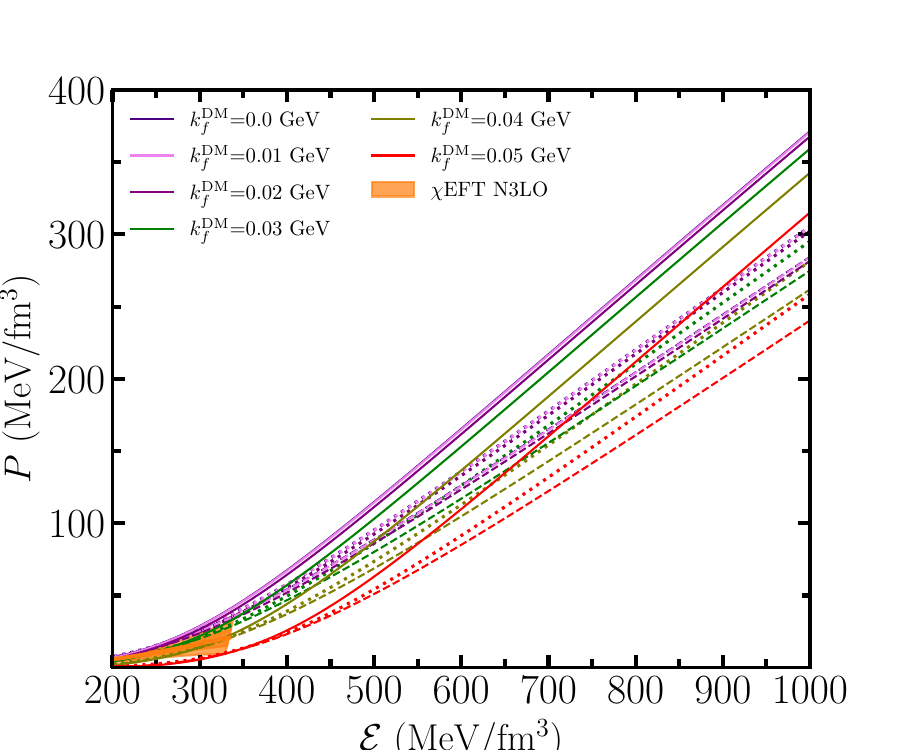}
    \includegraphics[width = 0.5\textwidth]{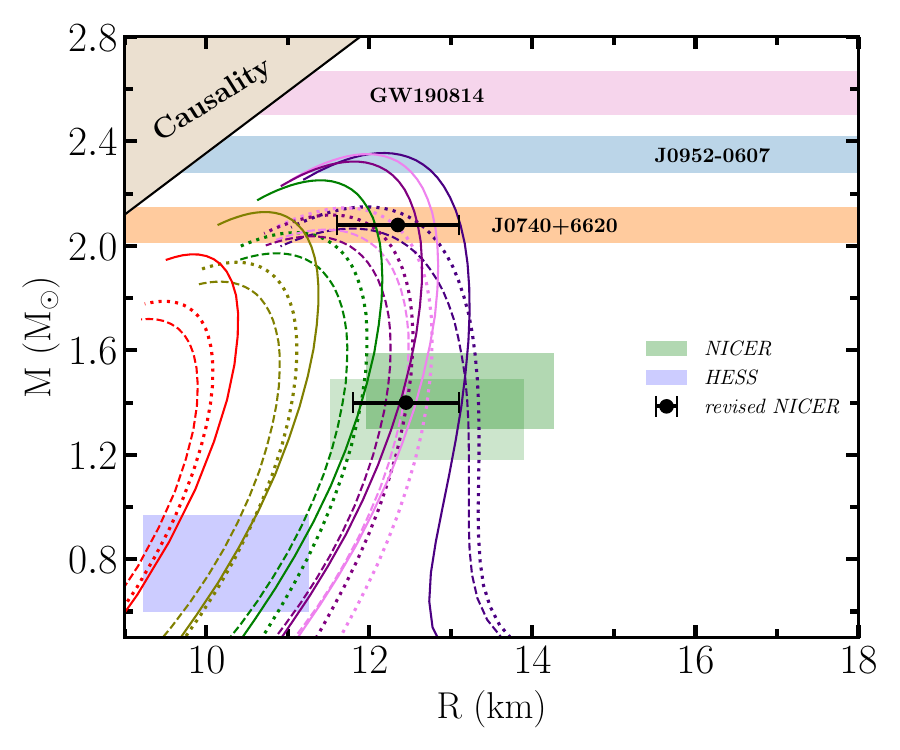}
    \includegraphics[width = 0.5\textwidth]{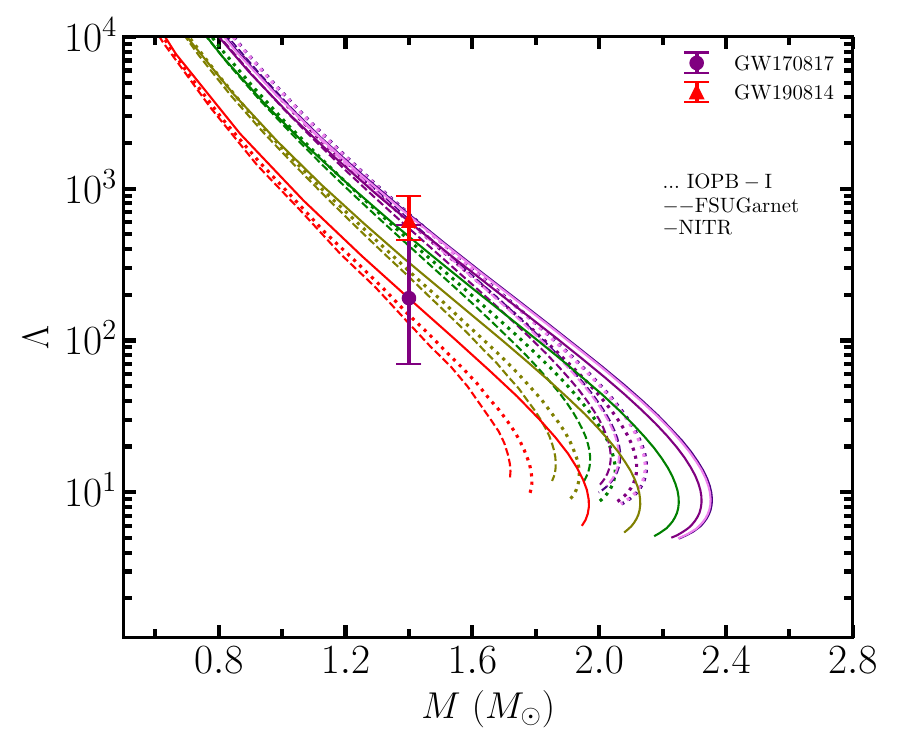}
    \caption{{\it Upper:} Unified EOSs for DMANS with and without DM by varying the $k_f^{\rm DM}$. {\it Lower left:} Mass-Radius relation for single fluid approach by varying $k_f^{\rm DM}$. Different color bands signify the masses of the NS observed from the various pulsars, such as PSR J0740+6620 \cite{Fonseca_2021}, heaviest pulsars J0952-0607 \cite{Romani_2022}, and GW190814 \cite{Abbott_2020}. The simultaneous mass-radius constraint provided by NICER \cite{Miller_2019, Riley_2019} and revised NICER \cite{Miller_2021} are also shown. {\it Lower right:} The tidal deformability as the function of the mass of the NS with varying $k_f^{\rm DM}$. Two observational constraints GW190814 \cite{Abbott_2020} and GW170817 \cite{Abbott_2017} are also used to validate our model.}
    \label{fig:smr-td}
\end{figure}

{In the case of SFDM, DM particles interact directly with nucleons through Higgs exchange, leading to the combined energy density and pressure of the system arising from both nucleons and DM. By varying $k_f^{\rm DM}$ within the IOPB-I, FSUGarnet, and NITR models, we compute the DMANS EOS. The unified EOS for DMANS is illustrated in the upper panel of Fig. \ref{fig:smr-td} for various $k_f^{\rm DM}$. Additionally, we superimpose the lower-density chiral EFT data \cite{Drischler_2021} on the graph. The inclusion of DM results in a softer EOS, and the extent of softness primarily depends on the amount of DM present within the NS.}

{The mass-radius relationship for our newly developed EOS `NITR', along with IOPB-I and FSUGarnet, is depicted in the lower left panel of Fig. \ref{fig:smr-td} with varying $k_f^{\rm DM}$. The heaviest pulsar, PSR J0952-0607 \cite{Romani_2022}, has a mass of $M = 2.35 \pm 0.17 \ M_\odot$, which is well-matched by our EOS, predicting a mass of $2.355 \ M_\odot$. NITR not only predicts the highest pulsar mass but also aligns with NICER \cite{Miller_2019, Riley_2019}, yielding a canonical radius of 13.13 km. However, with an increase in $k_f^{\rm DM}$, the maximum mass and corresponding radius decrease. This is attributed to the presence of DM, which softens the EOS, and this softening effect relies on the quantity of DM within the star. Observational data on mass and radius allow us to limit the quantity of DM within the NS. For $k_f^{\rm DM}$ up to 0.03 GeV, the behavior of the mass-radius relationship remains within observational boundaries and aligns with NICER, revised NICER as well as HESS data \cite{HESS_2022}.  Nonetheless, a further increase in $k_f^{\rm DM}$ leads to additional softening of the NS, causing the mass-radius behavior to deviate from observational constraints.}

 \begin{table}[tbp]
    \centering
    \caption{The canonical and maximum mass properties of the NS is shown for IOPB-I, FSUGarnet(FSUG) and NITR model with varying k$_f^{\rm DM}$.}
    \label{tab:sf_all_data}
    \renewcommand{\arraystretch}{1.5}
    \scalebox{0.7}{
    \begin{tabular}{ccccccccccccccccc}
    \hline\hline
    \multicolumn{1}{l}{\multirow{2}{*}{\begin{tabular}[c]{@{}l@{}}$k_f^{\rm DM}$\\(GeV)\end{tabular}}} & \multicolumn{1}{l}{\multirow{2}{*}{\begin{tabular}[c]{@{}l@{}} \hspace{0.5cm}Star\\ \hspace{0.5cm}type\end{tabular}}} &
    \multicolumn{3}{l}{\begin{tabular}[c]{@{}l@{}}\hspace{0.9cm}$M$\\\hspace{0.9cm}($M_\odot$)\end{tabular}}&
    \multicolumn{3}{l}{\begin{tabular}[c]{@{}l@{}}\hspace{0.9cm}$R$\\ \hspace{0.9cm}(km)\end{tabular}}& \multicolumn{3}{l}{\begin{tabular}[c]{@{}l@{}} \hspace{0.7cm}$\Lambda$\end{tabular}}& 
    \multicolumn{3}{l}{\begin{tabular}[c]{@{}l@{}} \hspace{0.7cm}$f$\\\hspace{0.7cm} (kHz)\end{tabular}}\\ 
    \cmidrule(lr){3-5}\cmidrule(lr){6-8}\cmidrule(lr){9-11} \cmidrule(lr){12-14}\cmidrule(lr){15-17}
    \multicolumn{1}{l}{} & \multicolumn{1}{l}{} &
    \multicolumn{1}{l}{IOPB-I} & \multicolumn{1}{l}{FSUG} & \multicolumn{1}{l}{NITR}&
    \multicolumn{1}{l}{\hspace{0.2cm}IOPB-I} & \multicolumn{1}{l}{\hspace{0.2cm}FSUG} & \multicolumn{1}{l}{NITR}&
    \multicolumn{1}{l}{\hspace{0.2cm}IOPB-I} & \multicolumn{1}{l}{\hspace{0.2cm}FSUG} & \multicolumn{1}{l}{NITR}&
    \multicolumn{1}{l}{\hspace{0.2cm}IOPB-I} & \multicolumn{1}{l}{\hspace{0.2cm}FSUG} & \multicolumn{1}{l}{NITR} \\ \hline
    \multicolumn{1}{l}{\multirow{2}{*}{}}0.00& \multicolumn{1}{l}{\hspace{0.5cm}Cano.}& 
    \multicolumn{1}{l}{1.400}    & \multicolumn{1}{l}{1.400}   & \multicolumn{1}{l}{1.400}&
    \multicolumn{1}{l}{13.33}    & \multicolumn{1}{l}{13.20}   & \multicolumn{1}{l}{13.13} &
    \multicolumn{1}{l}{685.81}    & \multicolumn{1}{l}{630.11}   & \multicolumn{1}{l}{682.84} & 
    \multicolumn{1}{l}{2.04}    & \multicolumn{1}{l}{2.09}   & \multicolumn{1}{l}{2.08}\\ 
    \multicolumn{1}{l}{} & \multicolumn{1}{l}{\hspace{0.5cm}Max.} &            
    \multicolumn{1}{l}{2.149}    & \multicolumn{1}{l}{2.066}   & \multicolumn{1}{l}{2.355} &
    \multicolumn{1}{l}{11.98}    & \multicolumn{1}{l}{11.77}   & \multicolumn{1}{l}{12.19}&
    \multicolumn{1}{l}{14.97}    & \multicolumn{1}{l}{16.97}   & \multicolumn{1}{l}{9.01}&
    \multicolumn{1}{l}{2.44}    & \multicolumn{1}{l}{2.50}   & \multicolumn{1}{l}{2.38}\\ \hline
    \multicolumn{1}{l}{\multirow{2}{*}{}}0.01 & \multicolumn{1}{l}{\hspace{0.5cm}Cano.} &
    \multicolumn{1}{l}{1.400}    & \multicolumn{1}{l}{1.400}   & \multicolumn{1}{l}{1.400}&
    \multicolumn{1}{l}{12.70}    & \multicolumn{1}{l}{12.41}   & \multicolumn{1}{l}{12.57}&
    \multicolumn{1}{l}{681.67}    & \multicolumn{1}{l}{617.84}   & \multicolumn{1}{l}{676.22}&
    \multicolumn{1}{l}{2.05}    & \multicolumn{1}{l}{2.11}   & \multicolumn{1}{l}{2.08}     \\ 
    \multicolumn{1}{l}{}  & \multicolumn{1}{l}{\hspace{0.5cm}Max.} &
    \multicolumn{1}{l}{2.145}   & \multicolumn{1}{l}{2.062}   & \multicolumn{1}{l}{2.351} &
    \multicolumn{1}{l}{11.74}    & \multicolumn{1}{l}{11.47}   & \multicolumn{1}{l}{11.99}&
    \multicolumn{1}{l}{15.07}    & \multicolumn{1}{l}{17.27}   & \multicolumn{1}{l}{8.99}&
    \multicolumn{1}{l}{2.45}    & \multicolumn{1}{l}{2.50}   & \multicolumn{1}{l}{2.38}     \\ \hline
    \multicolumn{1}{l}{\multirow{2}{*}{}} 0.02 & \multicolumn{1}{l}{\hspace{0.5cm}Cano.} & 
    \multicolumn{1}{l}{1.400}    & \multicolumn{1}{l}{1.400}   & \multicolumn{1}{l}{1.400} &
    \multicolumn{1}{l}{12.46}    & \multicolumn{1}{l}{12.20}   & \multicolumn{1}{l}{12.37}&
    \multicolumn{1}{l}{609.26}    & \multicolumn{1}{l}{553.93}   & \multicolumn{1}{l}{621.48}&
    \multicolumn{1}{l}{2.10}    & \multicolumn{1}{l}{2.16}   & \multicolumn{1}{l}{2.13}     \\  
    \multicolumn{1}{l}{} & \multicolumn{1}{l}{\hspace{0.5cm}Max.} &                             
    \multicolumn{1}{l}{2.118}    & \multicolumn{1}{l}{2.036}   & \multicolumn{1}{l}{2.322}     &
    \multicolumn{1}{l}{11.51}    & \multicolumn{1}{l}{11.33}   & \multicolumn{1}{l}{11.77} &
    \multicolumn{1}{l}{14.18}    & \multicolumn{1}{l}{17.70}   & \multicolumn{1}{l}{8.54}&
    \multicolumn{1}{l}{2.50}    & \multicolumn{1}{l}{2.53}   & \multicolumn{1}{l}{2.42}     \\ \hline
    \multicolumn{1}{l}{\multirow{2}{*}{}} 0.03 & \multicolumn{1}{l}{\hspace{0.5cm}Cano.}&
    \multicolumn{1}{l}{1.400}    & \multicolumn{1}{l}{1.400}   & \multicolumn{1}{l}{1.400}     &
    \multicolumn{1}{l}{11.89}    & \multicolumn{1}{l}{11.67}   & \multicolumn{1}{l}{11.90}&
    \multicolumn{1}{l}{466.90}    & \multicolumn{1}{l}{421.06}   & \multicolumn{1}{l}{497.69}&
    \multicolumn{1}{l}{2.23}    & \multicolumn{1}{l}{2.28}   & \multicolumn{1}{l}{2.23} \\  
    \multicolumn{1}{l}{} & \multicolumn{1}{l}{\hspace{0.5cm}Max.} &
    \multicolumn{1}{l}{2.050}    & \multicolumn{1}{l}{1.971}   & \multicolumn{1}{l}{2.251} &
    \multicolumn{1}{l}{11.08}    & \multicolumn{1}{l}{10.82}   & \multicolumn{1}{l}{11.73}&
    \multicolumn{1}{l}{13.85}    & \multicolumn{1}{l}{15.81}   & \multicolumn{1}{l}{8.49}&
    \multicolumn{1}{l}{2.59}    & \multicolumn{1}{l}{2.65}   & \multicolumn{1}{l}{2.51} \\ \hline \hline 
    \multicolumn{1}{l}{\multirow{2}{*}{}} 0.04 & \multicolumn{1}{l}{\hspace{0.5cm}Cano.}&
    \multicolumn{1}{l}{1.400}    & \multicolumn{1}{l}{1.400}   & \multicolumn{1}{l}{1.400}     &
    \multicolumn{1}{l}{11.04}    & \multicolumn{1}{l}{10.86}   & \multicolumn{1}{l}{11.15}&
    \multicolumn{1}{l}{291.05}    & \multicolumn{1}{l}{263.31}   & \multicolumn{1}{l}{328.95}&
    \multicolumn{1}{l}{2.44}    & \multicolumn{1}{l}{2.49}   & \multicolumn{1}{l}{2.42} \\  
    \multicolumn{1}{l}{} & \multicolumn{1}{l}{\hspace{0.5cm}Max.} &
    \multicolumn{1}{l}{1.937}    & \multicolumn{1}{l}{1.862}   & \multicolumn{1}{l}{2.129} &
    \multicolumn{1}{l}{10.36}    & \multicolumn{1}{l}{10.18}   & \multicolumn{1}{l}{10.68}&
    \multicolumn{1}{l}{12.94}    & \multicolumn{1}{l}{15.60}   & \multicolumn{1}{l}{8.23}&
    \multicolumn{1}{l}{2.76}    & \multicolumn{1}{l}{2.81}   & \multicolumn{1}{l}{2.67} \\ \hline \hline 
    \multicolumn{1}{l}{\multirow{2}{*}{}} 0.05 & \multicolumn{1}{l}{\hspace{0.5cm}Cano.}&
    \multicolumn{1}{l}{1.400}    & \multicolumn{1}{l}{1.400}   & \multicolumn{1}{l}{1.400}     &
    \multicolumn{1}{l}{10.05}    & \multicolumn{1}{l}{9.88}   & \multicolumn{1}{l}{10.24}&
    \multicolumn{1}{l}{147.75}    & \multicolumn{1}{l}{130.55}   & \multicolumn{1}{l}{192.91}&
    \multicolumn{1}{l}{2.73}    & \multicolumn{1}{l}{2.78}   & \multicolumn{1}{l}{2.69} \\  
    \multicolumn{1}{l}{} & \multicolumn{1}{l}{\hspace{0.5cm}Max.} &
    \multicolumn{1}{l}{1.787}    & \multicolumn{1}{l}{1.720}   & \multicolumn{1}{l}{1.967} &
    \multicolumn{1}{l}{9.450}    & \multicolumn{1}{l}{9.295}   & \multicolumn{1}{l}{9.810}&
    \multicolumn{1}{l}{11.42}    & \multicolumn{1}{l}{13.73}   & \multicolumn{1}{l}{8.00}&
    \multicolumn{1}{l}{3.02}    & \multicolumn{1}{l}{3.07}   & \multicolumn{1}{l}{2.90} \\ \hline \hline 
    \end{tabular}}
    \end{table}

\begin{figure}
   \centering
   \includegraphics[width = 0.6\textwidth]{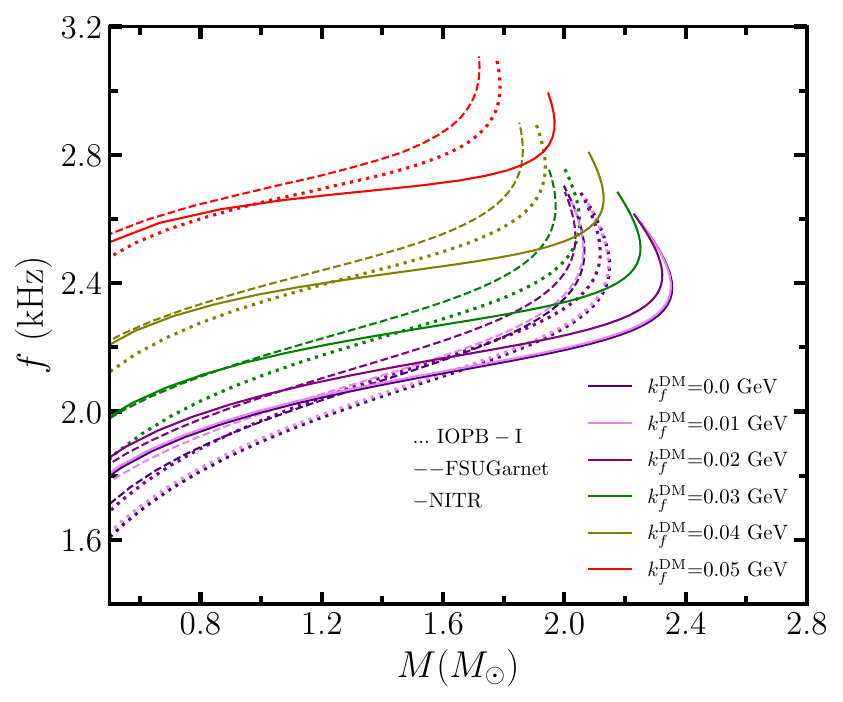}
    \caption{{The nonradial $f$-mode frequency as a function of mass is plotted while varying the k$_f^{\rm DM}$.}}
    \label{fig:fmode}
\end{figure}

The deformation of a NS in the presence of a companion star is measured by the tidal deformability, denoted by $\lambda=(2/3)k_2/C^5$, where $k_2$ and $C$ are the second Love number and compactness of the star \cite{Hinderer_2008, Kumartidal_2017} respectively. We calculate the dimensionless tidal deformability ($\Lambda=\lambda/M^5$) for the NITR EOS {along with IOPB-I and FSUGarnet} and display it in the lower right panel of Fig.\ref{fig:smr-td}. The predicted $\Lambda_{1.4}$ satisfies the limit ($\Lambda_{1.4}\leq 800$) given in Ref. \cite{Abbott_2017}; however, this limit is slightly above the upper bound of $\Lambda_{1.4}=190_{-70}^{+120}$ \cite{Abbott_2018}. The inclusion of DM inside the NS results in a significant change in the tidal deformability. When $k_f^{\rm DM}$ increases from 0.0 GeV to some finite value, the tidal deformability starts to decrease due to its dependence on the mass and radius of the NS. We increased $k_f^{\rm DM}$ up to 0.05 GeV and observed that the gravitational bound is satisfied for a canonical NS.
\begin{table}
\centering
\caption{{The constraint of canonical nonradial $f$-mode frequency by Wen $et \ al.$ \cite{Wen-fmode_2019}, Das {\it et al.} \cite{harishprd_2021}, and Sotani {\it et al.} are given.}}
\renewcommand{\tabcolsep}{0.15cm}
\renewcommand{\arraystretch}{1.5}
\begin{tabular}{cc}
\hline \hline
Previous Studies & $f_{1.4}$ \\
 \hline
Wen $et \ al.$ \cite{Wen-fmode_2019} & $1.67-2.18$ kHz \\
Das {\it et al.} \cite{harishprd_2021} & $1.78-2.22$ kHz \\
Sotani $et \ al.$ \cite{Sotani-kumar_2021}   & $1.68-2.57$ kHz \\ 
\hline \hline
\end{tabular}
\label{tab:comparision}
\end{table}

{In Fig. \ref{fig:fmode}, the behavior of the nonradial $f$-mode frequency is plotted against the NS mass while varying the parameter $k_f^{\rm DM}$. It has been observed that DM has significant impacts on $f$-mode frequency. The magnitude of $f$-mode frequency decreases with the increase in the $k_f^{\rm DM}$. The numerical values for all cases are provided in Table \ref{tab:sf_all_data}. Three different theoretical bounds are presented in Table \ref{tab:comparision}, where the constraints of the canonical nonradial $f$-mode frequency are enumerated. These constraints offer insights into constraining the amount of DM within the recently observed PSR J0952-0607. The canonical $f$-mode frequency for the NITR model is $f_{1.4}=2.08$ kHz, which is consistent with all the limit given in the Table \ref{tab:comparision}. As DM affects the frequency of the oscillation, the nonradial $f$-mode oscillation could provide an alternative method for constraining the quantity of DM within the NS. Analyzing both Table \ref{tab:sf_all_data} and Table \ref{tab:comparision}, it's evident that up to $k_f^{\rm DM} \approx$ 0.02, the NITR model aligns well with the constraint predicted by Wen {\it et al.} \cite{Wen-fmode_2019}, Sotani {\it et al.} \cite{Sotani-kumar_2021} and Das {\it et al.} \cite{harishprd_2021}.}

\subsection{Two fluid dark matter}
\begin{figure}
   \centering
    \includegraphics[width = 0.6\textwidth]{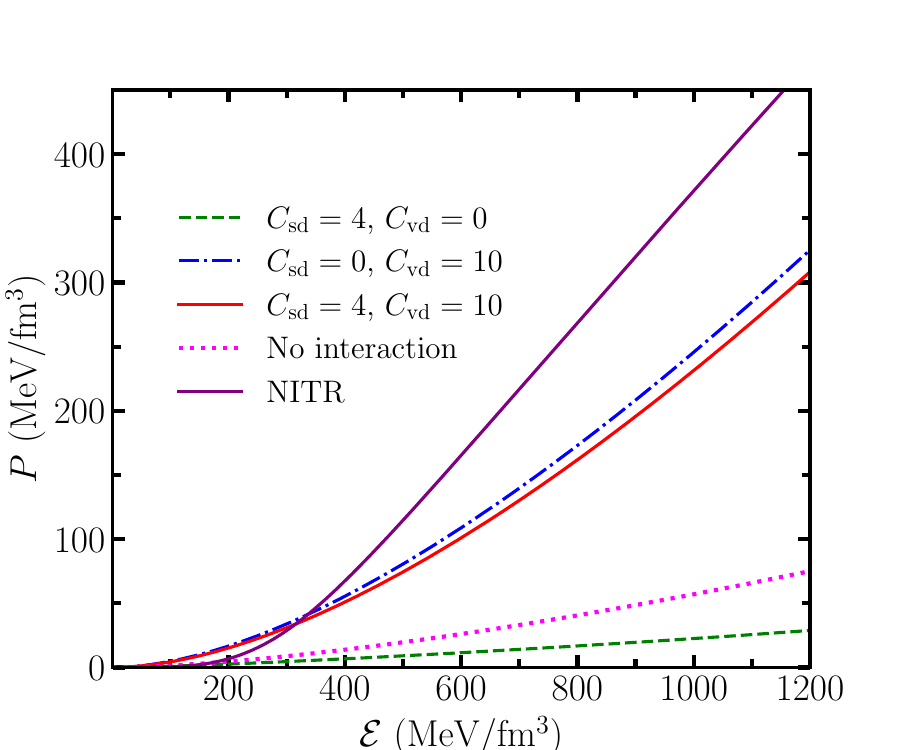}
    \caption{The TFDM EOSs for different choices of interaction strength or potential accounted here. Also, we take our new EOS `NITR' for comparison. The mass of the DM candidate is taken here as 1 GeV.}
    \label{fig:dm-eos}
\end{figure}

\begin{figure}
   \centering
    \includegraphics[width = 1.0\textwidth]{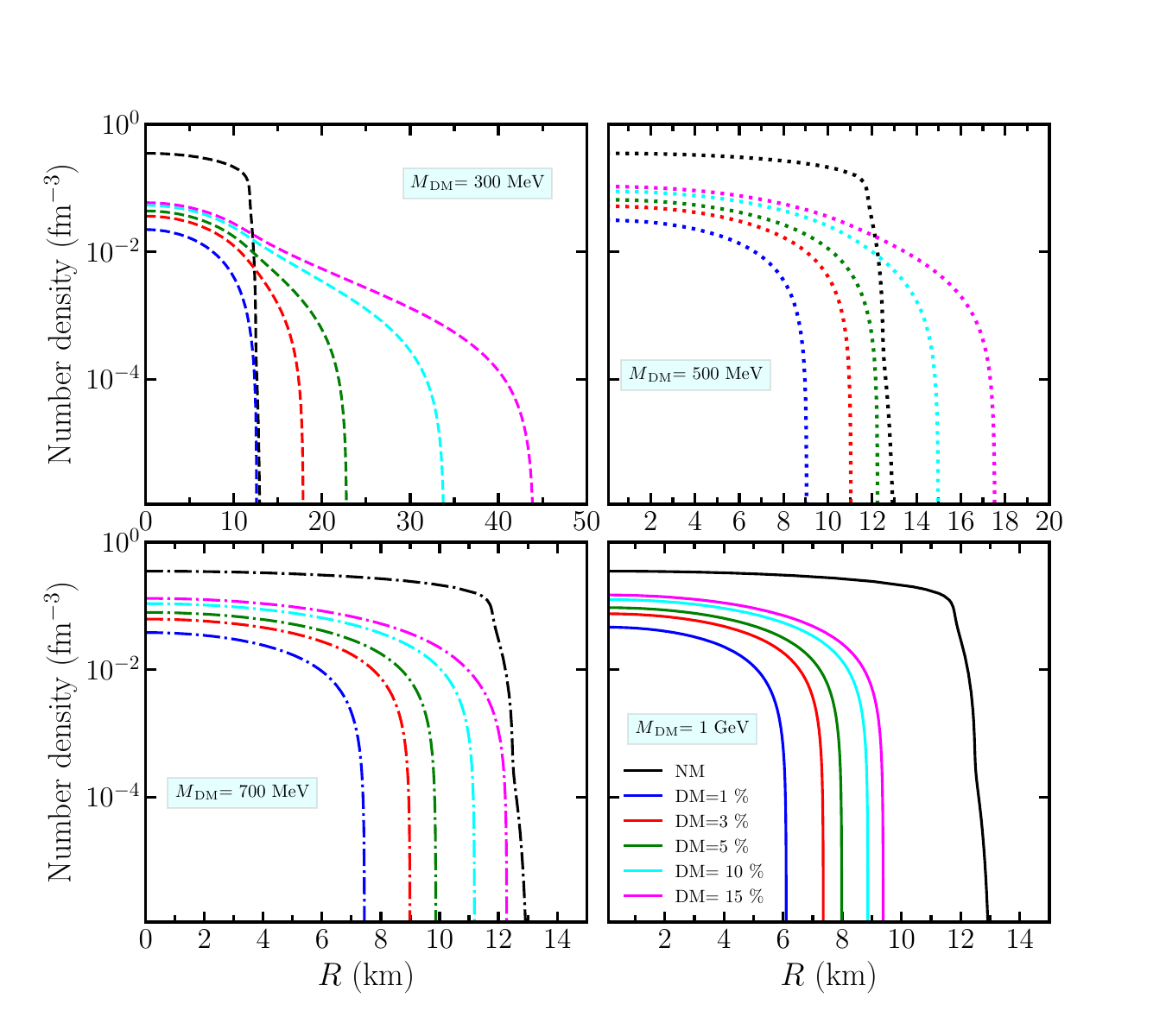}
    \caption{Density profile of the canonical NS for different DM masses by varying DM fraction. The possibility of DM halo and DM core is also shown. Here, the value of potential considered as $C_{\rm sd}=4$ GeV$^{-1}$ and $C_{\rm vd}=10$ GeV$^{-1}$.}
    \label{fig:prof}
\end{figure}

Here, we calculate the properties of DMANS, including its density profiles, $M-R$ relations, and tidal deformability, using the two-fluid model \cite{arpan_two-fluid_2022, Xiang_two-fluid_2014}. The predicted properties are obtained by varying the mass and fractions of DM inside the NS. We depict the TFDM EOSs in Fig. \ref{fig:dm-eos} for different potentials, including the EOS without DM for the NITR model, for comparison. The values of $C_{\rm sd}$ and $C_{\rm vd}$ are chosen such that they represent short-range repulsive and long-range attractive interactions, respectively. The exact values for these parameters are still unknown, but Ref. \cite{arpan_two-fluid_2022} predicts them using the Bayesian method and finds a range of C$_{\rm sd}$ = 3.90$^{\rm +0.82}_{\rm -0.70}$ GeV$^{-1}$ and C$_{\rm vd}$ = 11.88$^{\rm +0.53}_{\rm -0.46}$ GeV$^{-1}$, which are model-dependent. However, we roughly choose the values of C$_{\rm sd}$ = 0 and/or 4 GeV$^{-1}$ and C$_{\rm vd}$ = 0 and/or 10 GeV$^{-1}$ as done in \cite{Xiang_two-fluid_2014}. It is important to note that the smaller attractive potential slightly softens the DM EOS, whereas stiffening depends on the larger repulsive potential. However, the value of C$_{\rm sd}$ should not be too large as it can make the pressure negative and disrupt the hydrostatic equilibrium inside the star. Thus, we keep the values of C$_{\rm sd}$ and C$_{\rm vd}$ at 4 GeV$^{-1}$ and 10 GeV$^{-1}$, respectively, in our calculations.
We have obtained number density profiles as a function of the radius for the canonical star, which is depicted in Fig. \ref{fig:prof}, for different DM masses ranging from 300 MeV to 1 GeV and varying DM fractions. Our observations show that for lower DM masses, such as 300 MeV, an increase in the mass fraction results in an expansion of the DM radius, which leads to the formation of a DM halo that overpowers the NS radius. Conversely, for higher DM masses, such as 1 GeV, the radius of the NM dominates the DM halo, which is primarily concentrated at the center of the star. From this evidence, we can conclude that lighter DM particles tend to form DM halos, whereas heavier DM particles mainly reside in the core of DMANS \cite{Ivanytskyi_2020, Karkevandi_2022}.

\begin{figure}
   \centering
    \includegraphics[width = 0.7\textwidth]{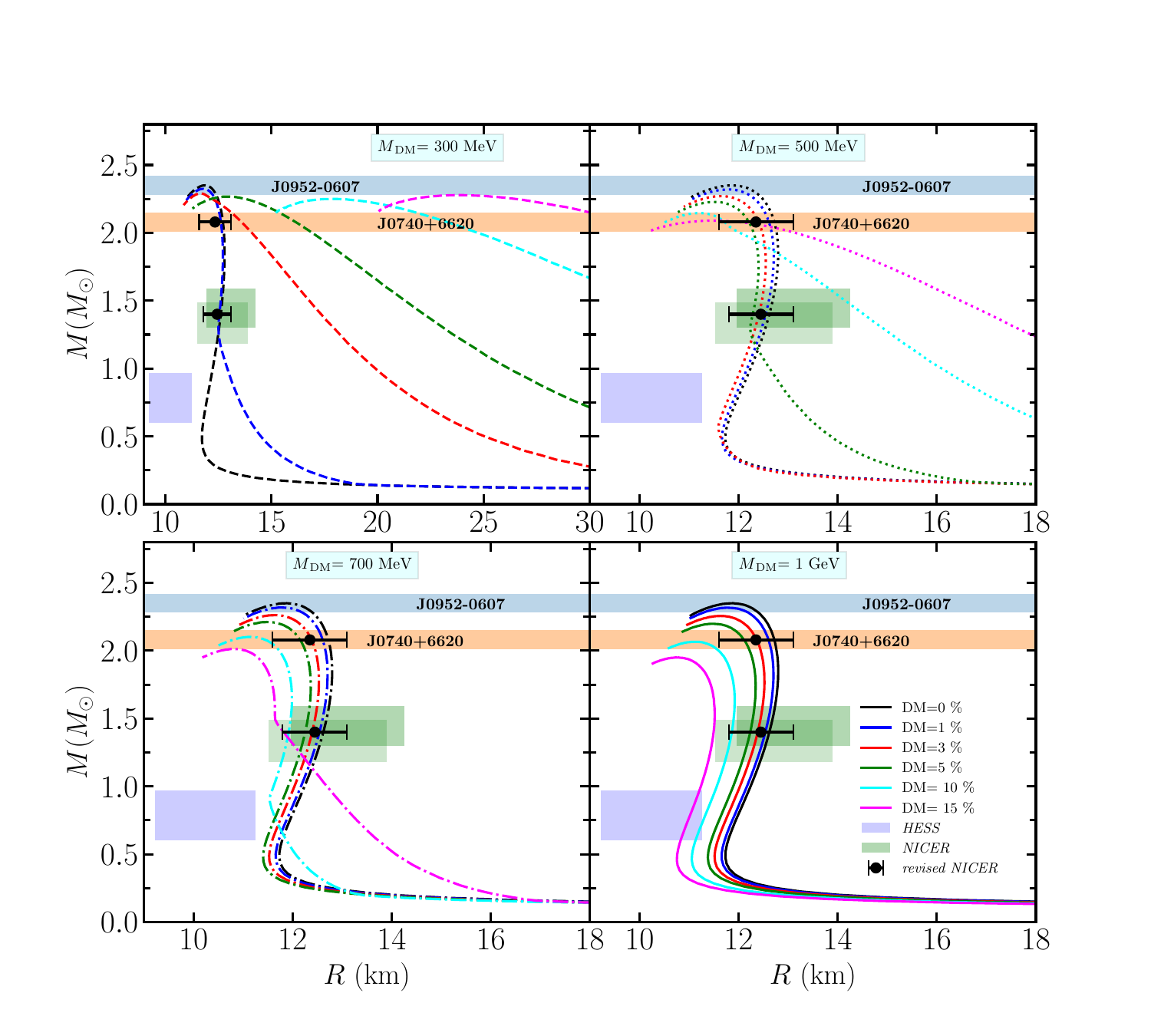}\\
    \includegraphics[width = 0.7\textwidth]{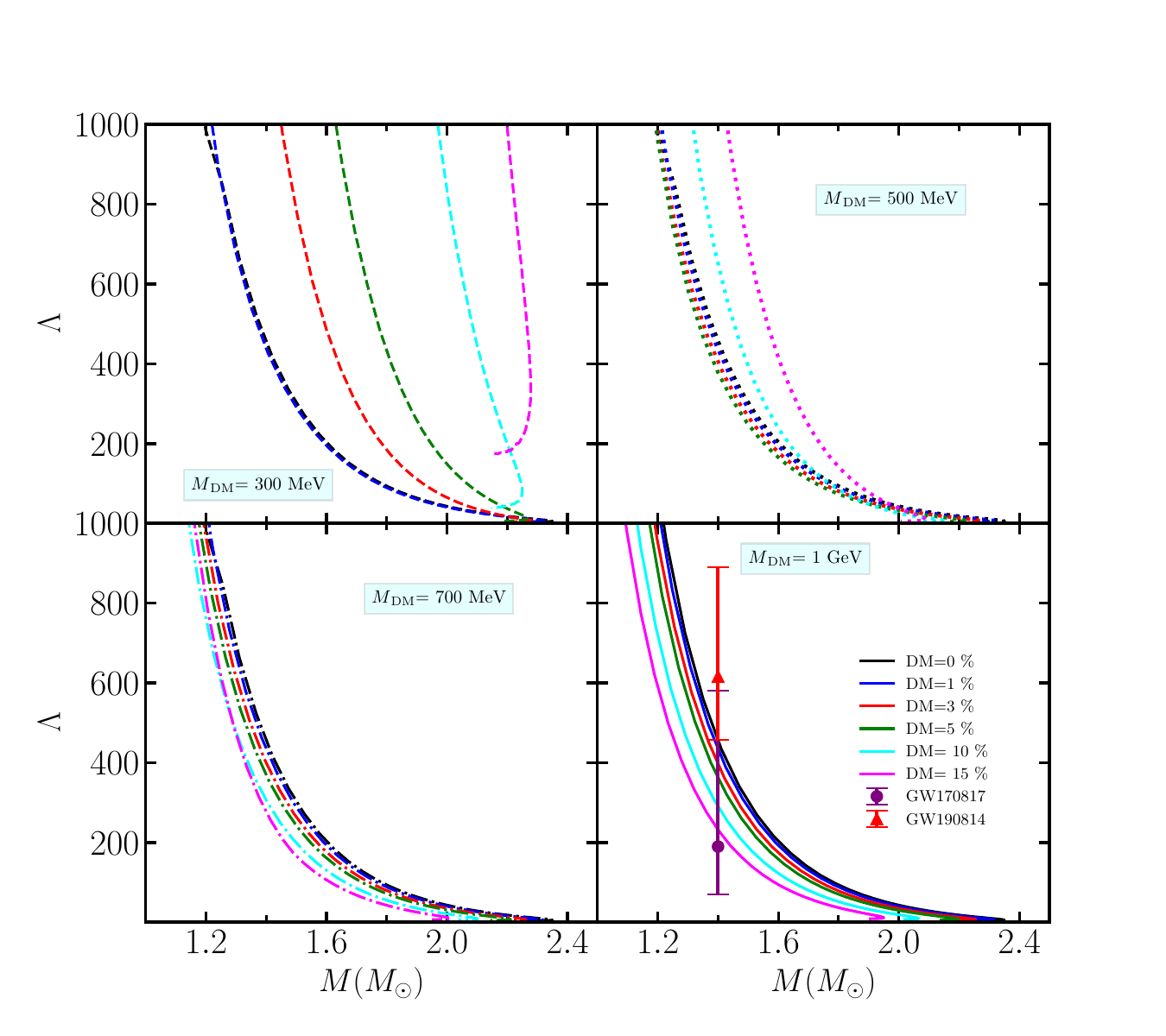}
    \caption{{\it Upper:} Mass-Radius relation for different DM particle mass ranging from 300 MeV to 1 GeV plotted varying mass fraction. The shaded bands represent different pulsars. {\it Lower:} The dimensionless tidal deformability is shown for the same condition as the upper panel.}
    \label{fig:mr}
\end{figure}

In Fig. \ref{fig:mr}, the upper panel shows the mass-radius relationship for different DM masses with varying mass fractions, taking into consideration different observational constraints to constrain the amount of dark matter present inside the NS. The orange shaded line represents the PSR J0740+6620 \cite{Miller_2021}, which suggests that any theoretical model's maximum mass limit should be able to estimate 2 $M_\odot$. In light of this, we vary the DM mass and mass fraction to constrain the amount of DM inside the NS while considering the impacts of DM on the maximum mass and radius of the NS. Interestingly, it is observed that the impacts of less massive DM candidates on the radius are more pronounced than those of heavier DM candidates. For example, when the DM mass is 300 MeV, an increase in the mass fraction leads to a continuous increase in the radius corresponding to the maximum mass, indicating the formation of a DM halo, as seen in Fig. \ref{fig:prof}. Similarly, for DM mass 500 MeV, a halo is observed for a higher mass fraction, while for a lower mass fraction, DM captures at the core of the NS. To observe this continuum, the DM mass is increased up to 1 GeV with varying mass fractions, where the DM completely captures inside the NS and resides at the core of the NS. Therefore, it can be concluded that less massive DM candidates dominate the NS to form a DM halo, whereas heavier DM candidates capture inside the NS and present as a DM core. {By considering the case of the large DM mass (1 GeV) in which DM is captured inside the core, the amount of DM can be constrained from both mass-radius constraint. Here up to $\sim$ 10\%, the mass-radius curve satisfying 2$M_\odot$ bound, radius constraint by revised NICER data \cite{Miller_2021} as well as mass-radius constraints by HESS J731$-$347.}

In the lower panel of Fig. \ref{fig:mr}, we present the predicted dimensionless tidal deformability. It is observed that the tidal deformability is dependent upon the radius of the star. When we increase the mass fraction of light-dark matter particles, the tidal deformability rises beyond the observational limits, as confirmed in Refs. \cite{arpan_two-fluid_2022}. For heavy DM particles such as 1 GeV, the tidal deformability decreases steadily with increasing mass fraction, leading to a canonical tidal deformability ($\Lambda_{1.4}$) that falls within the accepted limits of observation. In summary, the tidal deformability depends on the mass and mass fraction of the DM particles, and for heavier DM particles, the tidal deformability stays within the observed limits of the canonical tidal deformability.
\section{Conclusions}
In the present work, we have developed a new parameter set called `NITR' using the E-RMF formalism to support experimental evidence for finite nuclei, infinite nuclear matter, and neutron stars in light of the recently observed heaviest pulsar, PSR J0952-0607 ($M=2.35\pm0.17 \ M_\odot$). We optimized the new parameter set using empirical evidence for eight spherical nuclei, and the observables, such as binding energy and radii for finite nuclei, can now be predicted reasonably accurately. Our calculations are consistent with the existing experimental data for the infinite nuclear matter at saturation densities ($\rho_0=0.155$ fm$^{-3}$). We obtained incompressibility, symmetry energy, and slope parameter values for the NITR parameter set as 225.11, 31.69, and 43.86 MeV, respectively. The maximum mass yielded by our parameter set is consistent with PSR J0952-0607, and we obtained maximum mass of $2.355M_\odot$ and canonical radius in agreement with the theoretical results \cite{Nattila_2016, Annala_2018} with the value of 13.13 km. All the nuclear matter and NS properties of NITR model are compared with the well established models such as IOPB-I and FSUGarnet.

To investigate the NS macroscopic properties, we considered DMANS for our study, where we considered DM interacting with nucleons inside the NS both non-gravitationally (single fluid approach) and gravitationally (two fluid approach). In the single fluid approach, we considered neutralino as the DM candidate that interacts with nucleons by exchanging Higgs boson. By varying the $k_f^{\rm DM}$, we observed that the maximum mass and its corresponding radius, as well as canonical radius, decreases with an increase in $k_f^{\rm DM}$, and a similar effect was observed for tidal properties as they scale with mass and radius. Additionally, we have calculated the nonradial $f$-mode frequency for both with and without DM, and we have analyzed the impacts of DM on the amplitude of the oscillation. We constrained the amount of DM inside the NS using the canonical radius constraint based on revised NICER data, as well as the canonical nonradial $f$-mode frequency constraint from previous studies. We varied the value of $k_f^{\rm DM}$ in the range of $0.00$ to $0.05$ GeV. Our findings indicate that the properties of the NS satisfy observational constraints up to a specific value of $k_f^{\rm DM}$ (approximately $0.03$ GeV). This allows us to place constraints on the quantity of DM present inside the NS. 

On the other hand, in case of the two-fluid approach, we investigated the impacts of fermionic DM on the characteristics of the NS under the assumption that DM and NM interact only via gravity. First, we investigated the two-fluid EOS with different interaction strengths and finally choose their value in our computation in such a way that they should be short-range repulsive ($C_{\rm sd}$= 4 GeV$^{-1}$) and long-range attractive ($C_{\rm vd}$= 10 GeV$^{-1}$) interactions.
We defined multiple sets of mass of DM and mass fraction and took them as free parameters in our computations. We found that the mass-radius correlations of the NS and tidal characteristics both contract and broaden when DM is present. We established that the region of DM candidates that significantly affects the dynamics of nuclear structure is located around 300 MeV. Interestingly, when the mass of the DM is less, DM is not trapped inside the star and leads (can you please recheck this) to form the DM halo, which can be clearly understood from the density profile of the star. An explicit DM halo may be likely to develop if there is a repulsive force between DM particles. However, in the case of a heavier DM mass (1 GeV), when we varied the mass fractions, the DM became completely trapped inside the NS and settled inside the core. Consequently, while exploring various mass fractions, we discovered that the mass-radius curve remains consistent with the $2 M_\odot$ mass constraint and the radius constraint based on the revised NICER data for mass fractions up to approximately $10\%$. In the future, more observational data can put stringent constraints on the DM particles' nature and their amount inside compact objects.
\section{Acknowledgments}
 B.K. acknowledges partial support from the Department of Science and Technology, Government of India, with grant no. CRG/2021/000101.
\bibliographystyle{JHEP}
\bibliography{nitr.bib}
\end{document}